\documentclass{article}
\def\tr{{\rm tr}}

\begin{document}

\title{Dynamics of solutions of the Einstein equations with twisted Gowdy 
symmetry}

\author{Alan D. Rendall \\ 
Max-Planck-Institut f\"ur Gravitationsphysik\\
Albert-Einstein-Institut\\
Am M\"uhlenberg 1\\
14476 Potsdam, Germany}

\date{}

\maketitle

\begin{abstract}
Some of the most interesting results on the global dynamics of solutions
of the vacuum Einstein equations concern the Gowdy spacetimes whose 
spatial topology is that of a three-dimensional torus. In this paper 
certain of these ideas are extended to a wider class of vacuum spacetimes
where the spatial topology is that of a non-trivial torus bundle over a 
circle. Compared to the case of the torus these are topologically twisted.
They include inhomogeneous generalizations of the spatially homogeneous 
vacuum spacetimes of Bianchi types II and VI${}_0$. Using similar procedures
it is shown that the vacuum solutions of Bianchi type VII${}_0$ are
isometric to a class of Gowdy spacetimes, the circular loop spacetimes,
thus establishing links between results in the literature which were not 
previously known to be related to each other.

\end{abstract}

\section{Introduction}\label{intro}

The Gowdy spacetimes are a class of solutions of the vacuum Einstein equations
defined by certain symmetries. They are invariant under an action of the torus
$T^2$ and, in addition, they possess a discrete symmetry. In what follows only 
those Gowdy spacetimes are considered which have a compact Cauchy surface and 
are such that the transformations belonging to the action of $T^2$ have no 
fixed points. It can be concluded that the Cauchy surface is diffeomorphic
to the three-dimensional torus $T^3$ \cite{chrusciel90}. The discrete symmetry 
characterizes these spacetimes among the more general class of $T^2$-symmetric 
vacuum spacetimes whose dynamics is much less well understood. This 
characterization is explained in \cite{nungesser}. If it is assumed that in a 
solution of the Einstein equations coupled to matter the metric and the matter 
fields have the type of symmetry just described then they are said to have 
Gowdy symmetry. This paper is concerned with vacuum spacetimes but many of its 
results also apply to solutions with matter. The significance of the Gowdy 
solutions on $T^3$ for the more general task of investigating the dynamics of 
solutions of the vacuum Einstein equations is that they represent the simplest 
class of inhomogeneous spatially compact solutions of these equations. As 
such they are an ideal laboratory for studying certain phenomena.

The subject of this paper is a generalization of Gowdy symmetry which may be 
called twisted Gowdy symmetry (as it is in this paper) or local Gowdy 
symmetry. The former name is related to the fact that in many cases the metric 
is naturally defined on a manifold which is topologically twisted. (This has 
nothing to do with the 'twist constants' whose vanishing is often used as a 
characterization of Gowdy spacetimes among $T^2$-symmetric spacetimes, cf. 
\cite{rendall08}, section 4.4.) Spacetimes which generalize those with Gowdy 
symmetry in the way considered in what follows have previously been discussed 
in \cite{rendall97}, \cite{tanimoto98}, \cite{tanimoto00}, \cite{tanimoto01} 
and \cite{weaver98}. Analogous symmetry assumptions have been studied for
the Ricci flow in \cite{hamilton93}. In many cases 
the group $T^2$ does not act on the solution itself but only on its pull-back 
to the universal covering manifold. The question of central interest in what 
follows is the dynamics of these solutions. There are two asymptotic regimes, 
the approach to the initial singularity and the late-time 
behaviour. The dynamics near the initial singularity in Gowdy spacetimes is 
very well understood \cite{ringstrom06a}, \cite{ringstromta}. Many of the 
arguments used in that work are local in space and so can be applied to get 
analogous conclusions in the twisted Gowdy case. There are, nevertheless, 
important questions in this context which are still open. This is discussed at 
the end of the next section.  In the case of the late-time behaviour there is 
no reason to expect spatial localization. This paper focusses on the late-time 
behaviour, examining to what extent it differs from the known facts about the 
Gowdy case obtained in \cite{ringstrom04}.

The second section introduces the basic definitions and equations for 
twisted Gowdy spacetimes and explains how they can be regarded as 
generalizations of Bianchi models of types II and VI${}_0$, just as
the ordinary Gowdy spacetimes are generalizations of Bianchi type I.
Section \ref{bianchi7} discusses a topic related to the main theme of this
paper, establishing a link between the circular loop Gowdy solutions and 
spacetimes of Bianchi type VII${}_0$. Section \ref{latetime} discusses the 
late-time asymptotic behaviour of the twisted models. In section 
\ref{linearized} the nonlinear analysis done in this paper is compared to 
existing work on linear perturbations of Bianchi models. The last section
presents some conclusions and an outlook on possible future developments.

\section{Basic equations}\label{basic}

In the Gowdy class the spacetime metric can be written in the form
\begin{equation}\label{gowdymetric}
t^{-\frac12}e^{\frac{\lambda}2}(-dt^2+d\theta^2)+t(e^P(dx+Qdy)^2+e^{-P}dy^2).
\end{equation}
Here the functions $P$, $Q$ and $\lambda$ depend only on $t$ and $\theta$.
The essential equations describing these spacetimes are a system of semilinear 
wave equations for two functions $P$ and $Q$ which are assumed to be periodic 
with period $2\pi$ in the spatial coordinate $\theta$. The equations are
\begin{eqnarray}
&&P_{tt}+t^{-1}P_t=P_{\theta\theta}+e^{2P}
(Q_t^2-Q_\theta^2)\label{gowdyP},\\
&&Q_{tt}+t^{-1}Q_t=Q_{\theta\theta}-2(P_tQ_t-P_\theta Q_\theta )\label{gowdyQ}
\end{eqnarray}
where the subscripts denote partial derivatives. In what follows we call
(\ref{gowdyP}) and  (\ref{gowdyQ}) the Gowdy equations.

It is useful to interpret these equations as a wave map, a concept which 
will now be recalled. Let $(M,g)$ be a pseudo-Riemannian and $(N,h)$ a 
Riemannian manifold and let $\Phi$ be a smooth mapping from $M$ to $N$.
Define a Lagrangian density by $L=\frac{\partial\Phi^I}{\partial x^\alpha}
\frac{\partial\Phi^J}{\partial x^\beta}g^{\alpha\beta}h_{IJ}$. The
corresponding Euler-Lagrange equations define what are called harmonic maps 
when $g$ is Riemannian and wave maps when $g$ is Lorentzian. The Riemannian
manifold $(N,h)$ is referred to as the target space. 

The equations (\ref{gowdyP}) and (\ref{gowdyQ}) can be interpreted as the 
defining equations of wave maps from the auxiliary metric
\begin{equation}\label{auxiliary}
-dt^2+d\theta^2+t^2 d\phi^2
\end{equation}
to the hyperbolic plane which are independent of the coordinate $\phi$. In
this context $P$ and $Q$ are interpreted as coordinates on the hyperbolic plane
which put its metric into the form $dP^2+e^{2P}dQ^2$. The remaining function 
$\lambda$ in the spacetime metric is given by integrals where the 
integrands are determined by $P$ and $Q$. This follows from the equations
\begin{eqnarray}
&&\lambda_t=t[P_t^2+P_\theta^2+e^{2P}(Q_t^2+Q_\theta^2)],\label{lambdat} \\
&&\lambda_\theta=2t(P_tP_\theta+e^{2P}Q_tQ_\theta)\label{lambdatheta}.
\end{eqnarray}
To ensure that the metric coefficient $\lambda$ is periodic it is necessary 
to require the condition
\begin{equation}\label{lambdaint}
\int_0^{2\pi}P_tP_\theta+e^{2P}Q_tQ_\theta d\theta=0.
\end{equation}
If this condition is satisfied at one time it is satisfied at all times,
since 
\begin{equation}
\frac{d}{dt}\left(\int_0^{2\pi}P_tP_\theta+e^{2P}Q_tQ_\theta d\theta\right)
=-\frac{1}{t}\left(\int_0^{2\pi}P_tP_\theta+e^{2P}Q_tQ_\theta d\theta\right).
\end{equation} 

In Gowdy spacetimes $P$ and $Q$ are periodic in $\theta$ with period $2\pi$. 
The aim of this paper is to study the dynamics of solutions of equations 
(\ref{gowdyP}) and (\ref{gowdyQ}) with other types of boundary 
conditions. Another way of describing the periodic boundary conditions is 
to say that the pull-back of the solution to the universal cover is invariant
under a translation in $\theta$ by $2\pi$. The more general boundary 
conditions correspond to replacing invariance under a translation by $2\pi$ 
in $\theta$ by equivariance under that translation. Let $X$ be a Killing 
vector of the hyperbolic plane and $\Psi_\gamma$ the one-parameter family of 
diffeomorphisms it generates. Equivariant wave maps are defined by the 
condition that 
\begin{equation}
(P,Q)(t,\theta+\gamma)=\Psi_\gamma ((P,Q)(t,\theta)).
\end{equation}
In what follows an equivariant wave map is defined to be one which satisfies
this condition for all real numbers $\gamma$. The restriction of this to
integer multiples of $2\pi$ defines the generalizations of periodic boundary 
conditions to be considered. The solutions of the Gowdy equations with these
boundary conditions can be used to construct spacetimes defined on manifolds 
corresponding to Bianchi types II and VI${}_0$ which are bundles over the 
circle $S^1$ whose fibre is the torus $T^2$. These solutions contain a 
topological twist and for this reason will be said to have twisted Gowdy 
symmetry. They are inhomogeneous generalizations of the corresponding Bianchi 
models. 

Let the initial data for $P$, $P_t$, $Q$ and $Q_t$ on some
hypersurface $t=t_0$ be denoted by 
$P_0$, $P_1$, $Q_0$ and $Q_1$ respectively. Consider data for which $P_0$, 
$P_1$ and $Q_1$ are periodic while $Q_0(\theta+2\pi)=Q_0(\theta)+2\pi\alpha$ 
for a constant $\alpha\ne 0$. This type of data will be referred to as twisted 
data of type II. It is related to equivariance with respect to the Killing
vector $\alpha\frac{\partial}{\partial Q}$. It is known that any periodic 
data for the Gowdy equations prescribed at some time $t=t_0>0$ give 
rise to a unique corresponding global solution on the time interval 
$(0,\infty)$ \cite{moncrief81}. 
Moreover, as a result of the domain of dependence, the solution 
at a point with coordinates $(t,\theta)$ is uniquely determined by the 
data on the interval $[\theta_1,\theta_2]$ of the initial hypersurface $t=t_0$,
where $\theta_1=\theta-|t-t_0|$ and $\theta_2=\theta+|t-t_0|$. From
this global existence and uniqueness can be concluded for solutions of the 
Gowdy equations corresponding to data on the real line without imposing any 
spatial boundary conditions. It suffices to piece together suitable solutions
which are defined locally in space. Using uniqueness it can be shown that the 
solution corresponding to twisted data of type II is such that $P$ is periodic 
in $\theta$ while $Q(t,\theta+2\pi)=Q(t,\theta)+2\pi\alpha$. For in this
case  $(P(t,\theta),Q(t,\theta))$ and $(P(t,\theta+2\pi),
Q(t,\theta+2\pi)-2\pi\alpha)$ 
are solutions of the Gowdy equations with the same initial data and hence must 
be equal. Note that the condition (\ref{lambdaint}) generalizes in an obvious 
way to twisted data of type II since the integrand occurring there is periodic.
When this condition is satisfied it is possible to find a solution $\lambda$
of the equations (\ref{lambdat}) and (\ref{lambdatheta}) which is periodic in 
$\theta$ since the right hand side of (\ref{lambdat}) is periodic. The
metric which has been given in local coordinates defines a spatially compact
spacetime provided $\pi\alpha$ is an integer. 

Next consider data 
satisfying $P_0(\theta+2\pi)=P_0(\theta)+2\pi\alpha$ for a constant 
$\alpha\ne 0$, $P_1(\theta+2\pi)=P_1(\theta)$, 
$Q_0(\theta+2\pi)=e^{-2\pi\alpha}Q_0(\theta)$,
$Q_1(\theta+2\pi)=e^{-2\pi\alpha}Q_1(\theta)$. This will be 
referred to as twisted data of type VI${}_0$. It is related to equivariance 
with respect to the Killing vector $\alpha(\frac{\partial}{\partial P}-
Q\frac{\partial}{\partial Q})$. Using the same methods as in the type II case 
it can be shown that there exist unique global solutions corresponding to data 
of this kind and that these solutions satisfy the conditions that 
$P(t,\theta+2\pi)=P(t,\theta)+2\pi\alpha$ and 
$Q(t,\theta+2\pi)=e^{-2\pi\alpha}Q(t,\theta)$.
As in type II the integrand in (\ref{lambdaint}) is periodic for these 
solutions and so the restriction which ensures the periodicity of $\lambda$
is well-defined. The metric which has been given in local coordinates defines 
a spatially compact spacetime provided $\pi\alpha$ is an eigenvalue of a
matrix in $GL(2,Z)$. (For the significance of $GL(2,Z)$ in this context see
section 2 of \cite{rendall97}.) In the type VI${}_0$ case there is a class of 
polarized solutions defined by the vanishing of $Q$. Note for comparison that 
the boundary conditions for twisted Gowdy solutions of type II are not 
consistent with setting $Q=0$.

The terminology involving type II and type VI${}_0$ is explained by the fact 
that there are special solutions satisfying these boundary conditions which 
correspond to spatially homogeneous solutions of the respective Bianchi types. 
The first solution is obtained by setting 
$P(t,\theta)=\bar P(t)$ and $Q(t,\theta)=\alpha\theta$. 
The metric takes the form
\begin{equation}\label{bianchiII}
t^{-\frac12}e^{\frac{\lambda}{2}}(-dt^2+d\theta^2)+
t\left[e^{\bar P(t)}(dx+\alpha \theta dy)^2+e^{-\bar P(t)}dy^2\right].
\end{equation}
With these assumptions equation (\ref{gowdyQ}) is satisfied automatically 
while (\ref{gowdyP}) reduces to
\begin{equation}\label{gowdyPhom}
\frac{d^2\bar P}{dt^2}+t^{-1}\frac{d\bar P}{dt}=-\alpha^2e^{2\bar P}.
\end{equation}
Condition (\ref{lambdaint}) is also satisfied because the integrand vanishes. 
The metric (\ref{bianchiII}) is a spatially homogeneous spacetime which is 
expressed in terms of the one-forms $d\theta$, $dx+\alpha \theta dy$ and $dy$ 
with dual basis 
$\frac{\partial}{\partial\theta}$, 
$\frac{\partial}{\partial x}$, 
$\frac{\partial}{\partial y}-\alpha \theta\frac{\partial}{\partial x}$.
Computing the commutators of these vector fields reveals that this is a
metric of Bianchi type II. This metric corresponds directly to the expression 
for the Taub solutions on p. 196 of \cite{wainwright}. The latter is given by
\begin{equation}\label{taub}
-A^2dt^2+t^{2p_1}A^{-2}(dx+4p_1bzdy)^2+t^{2p_2}A^2dy^2+t^{2p_3}A^2dz^2
\end{equation}
where $p_1$, $p_2$ and $p_3$ are constants satisfying the Kasner relations
$p_1+p_2+p_3=1$ and $p_1^2+p_2^2+p_3^2=1$ while $A^2=1+b^2t^{4p_1}$. In 
order that the solution be of Bianchi type II it is important that $p_1\ne 0$
and $b\ne 0$. Otherwise a Bianchi I solution is obtained. When $p_1=0$ it
is the flat Kasner solution. When $p_1\ne 0$ it follows that $p_3< 1$ and 
the metric (\ref{taub}) can be put into the Gowdy form (\ref{gowdymetric}) by 
defining 
$\tilde t=(1-p_3)^{-1}t^{1-p_3}$, $\theta=z$, $\tilde x=(1-p_3)^{\frac12}x$,
$\tilde y=(1-p_3)^{\frac12}y$, $\alpha=4p_1b$, $P=(p_1-p_2)\log t-2\log A$ 
and $\lambda=(3p_3+1)\log t+4\log A$. Here $A$, $P$ and $\lambda$ should be 
thought of as functions of $\tilde t$ via the relation 
$t=[(1-p_3)\tilde t]^{\frac{1}{1-p_3}}$. 
These are the most general vacuum solutions of Bianchi type II. More 
precisely, the pull-back of any vacuum solution of Bianchi type II to the 
universal covering manifold can be written globally in the form given in 
(\ref{taub}). 

Another type of solution of equations (\ref{gowdyP}) and (\ref{gowdyQ}) is 
obtained by setting $P(t,\theta)=\bar P(t)+\alpha\theta$, 
$Q(t,\theta)=e^{-\alpha\theta}\bar Q(t)$. 
The metric on the group orbits is then of the form
\begin{equation}\label{VIPQ}
e^{\bar P(t)}(e^{\frac12\alpha\theta}dx+\bar Q(t)e^{-\frac12\alpha\theta}dy)^2
+e^{-\bar P(t)}(e^{-\frac12\alpha\theta}dy)^2.
\end{equation}
Equations (\ref{gowdyP}) and (\ref{gowdyQ}) become 
\begin{eqnarray}
&&\frac{d^2\bar P}{dt^2}+t^{-1}\frac{d\bar P}{dt}=e^{2\bar P}
\left[\left(\frac{d\bar Q}{dt}\right)^2-\alpha^2\bar Q^2\right],
\\
&&\frac{d^2\bar Q}{dt^2}+t^{-1}\frac{d\bar Q}{dt}=-2
\frac{d\bar P}{dt}\frac{d\bar Q}{dt}-\alpha^2\bar Q.
\end{eqnarray}  
Provided the right hand side of (\ref{lambdatheta}) vanishes a spatially 
homogeneous spacetime is obtained expressed in terms of the one-forms 
$d\theta$, 
$e^{\frac12\alpha\theta}dx$ and 
$e^{-\frac12\alpha\theta} dy$ with dual basis 
$\frac{\partial}{\partial\theta}$, 
$e^{-\frac12\alpha\theta}\frac{\partial}{\partial x}$,
$e^{\frac12\alpha\theta}\frac{\partial}{\partial y}$.
Computing the commutators of these vector fields shows that this is 
a solution of Bianchi type VI${}_0$. 

The vanishing condition for the right hand side of (\ref{lambdatheta}) and 
the evolution equations for $\bar P$ and $\bar Q$ are rather 
complicated. A more transparent formulation can be obtained by introducing 
the variables
\begin{eqnarray}
&&V=\log [e^{P}Q+\sqrt{1+e^{2P}Q^2}],     \\
&&W=-\frac12\log [e^{-2P}+Q^2].
\end{eqnarray}
The mapping $(P,Q)\mapsto (V,W)$ is smooth and has a smooth inverse which
is given explicitly as follows:
\begin{eqnarray}
&&P=W+\log\cosh V,   \\
&&Q=e^{-W}\tanh V.
\end{eqnarray}
In the variables $(V,W)$ the wave map equations take the form
\begin{eqnarray}
W_{tt}+t^{-1}W_t=W_{\theta\theta}+\tanh V(-W_tV_t+W_\theta V_\theta), \\
V_{tt}+t^{-1}V_t=V_{\theta\theta}-\cosh V\sinh V (-W_t^2+W_\theta^2).
\end{eqnarray}
When $P$ and $Q$ satisfy the type VI${}_0$ boundary conditions $V$ is 
periodic while $W(t,\theta+2\pi)=W(t,\theta)+2\pi\alpha$. 
The polarized class is characterized by $V=0$.
The restricted class of solutions transforms to solutions of the form
$V(t,\theta)=\bar V(t)$ and $W(t,\theta)=\bar W(t)+\alpha\theta$. They
are equivariant with respect to the vector field 
$\alpha\frac{\partial}{\partial W}$. For
this class the evolution equations for $V$ and $W$ reduce to
\begin{eqnarray}
&&\bar W_{tt}+t^{-1}\bar W_t=-\tanh\bar V\bar W_t\bar V_t\label{Wbar},\\
&&\bar V_{tt}+t^{-1}\bar V_t=-\cosh\bar V\sinh\bar V (-\bar W_t^2+\alpha^2).
\label{Vbar}
\end{eqnarray}
The condition for the periodicity of $\lambda$ simplifies to $\bar W_t=0$.
From now on it will be assumed that $\bar W=0$. Then (\ref{Vbar}) simplifies
to
\begin{equation}\label{Vbar2}
\bar V_{tt}+t^{-1}\bar V_t=-\alpha^2\cosh\bar V\sinh\bar V.
\end{equation}
The evolution equation for $\lambda$ is 
\begin{equation}
\lambda_t=t[\cosh^2 V(W_t^2+W_\theta^2)+(V_t^2+V_\theta^2)].
\end{equation}
The metric on the group orbits takes the form
\begin{equation}\label{VIVW}
\frac12 e^{\bar V} (e^{\frac12\alpha\theta}dx+e^{-\frac12\alpha\theta}dy)^2
+\frac12 e^{-\bar V} (e^{\frac12\alpha\theta}dx-e^{-\frac12\alpha\theta}dy)^2.
\end{equation}
Here the metric is diagonal in a left-invariant basis, in contrast to
the metric (\ref{VIPQ}). In the terminology of \cite{ringstrom09a} the basis 
used in (\ref{VIVW}) is canonical and the fact that the metric is diagonal
may be put into context by comparing with Corollary 19.14 of that reference
which says that any solution of Bianchi class A can be diagonalized in a 
canonical frame. The special case obtained by setting $\bar Q=0$ in the 
metric (\ref{VIPQ}) (polarized case) gives rise to a class of spacetimes 
which can be identified with the Ellis-MacCallum solutions given on p. 197 
of \cite{wainwright}. Note that in this case the condition (\ref{lambdaint}) 
forces $\bar P=0$.

A large class of twisted Gowdy solutions of Bianchi type II can be obtained 
starting from ordinary Gowdy solutions using the Gowdy-to-Ernst 
transformation. This transformation was introduced in the study of spikes
in Gowdy spacetimes \cite{rendall01} and was later used in the study of the 
initial singularity \cite{ringstrom06a}. The definition of the transformation
is as follows. Given a solution $(P,Q)$ of the Gowdy equations define a new
solution $(\tilde P,\tilde Q)$ by the relations
\begin{eqnarray}
&&\tilde P=-\log t-P,\\
&&\tilde Q_t=te^{2P}Q_\theta,\ \ \ \ \tilde Q_\theta=te^{2P}Q_t.
\end{eqnarray}
Determining $\tilde Q$ requires some integration and while this is always 
possible locally it is only possible globally on the torus if
\begin{equation}\label{geglobal}
\int_0^{2\pi}te^{2P}Q_t d\theta=0.
\end{equation}
If the integral in this equation has a suitable non-zero value then 
$\tilde P$ and $\tilde Q$ define a twisted Gowdy solution of type II. Note 
that 
\begin{equation}
\frac{d}{dt}\left(\int_0^{2\pi}te^{2P}Q_t d\theta\right)=0,
\end{equation}
so that it is enough to require the condition at one time in order
to ensure that it is satisfied at all times. The 
condition (\ref{lambdaint}) is preserved by this transformation and so if a 
periodic $\lambda$ exists before transformation the same is true after 
transformation. Not all solutions of type II are obtained in this way.
A necessary and sufficient condition for a solution to be contained in 
the image of this transformation is that it satisfies (\ref{geglobal}).

It will now be shown how certain statements about the behaviour of solutions
near the initial singularity can be transferred from the standard Gowdy case
to twisted Gowdy solutions. Consider a solution of the Gowdy equations 
defined on a region of the form $S=(0,t_1)\times I$ for a open interval $I$. 
Consider a point $(t_0,\theta_0)$ for which $t$ is so small that the interval 
$(\theta_0-2t_0,\theta_0+2t_0)$ is contained in $I$. Then the part of the 
solution in the past of $(t_0,\theta_0)$ is determined by data on the part of 
$t=t_0$ which is contained in $I$. Suppose in addition that the length of $I$ 
is less than $2\pi$. Then it is elementary to see that the solution on the 
past of $(t_0,\theta_0)$ can be embedded into a solution with periodic 
boundary conditions of period $2\pi$. Define the asymptotic velocity at
$\theta_0$ to be
\begin{equation}
\lim_{t\to 0} [t(P_t^2+e^{2P}Q_t^2)^{1/2}](t,\theta_0).
\end{equation} 
The asymptotic velocity defines a function of $\theta$ which is periodic.
It is proved in \cite{ringstrom06a} that in a Gowdy solution this limit exists
at any point $\theta_0$. From the remarks just made about embeddings it 
follows that the same conclusion holds for twisted Gowdy solutions. The 
notions of true and false spikes as defined in \cite{ringstromta} make 
sense for twisted Gowdy solutions. Thus as in that paper it is possible to
define the set ${\cal G}_c$ of twisted Gowdy solutions which satisfy
(\ref{lambdaint}), have non-degenerate true and false spikes at a finite number 
of values of $\theta$ and are such that the asymptotic velocity is strictly
between zero and one everywhere else. To define a notion of genericity it
is necessary to define a suitable topology on the set of twisted solutions 
of a given type. This can be done by using the standard $C^\infty$ topology on
an interval of length $2\pi$ and noting that the result does not depend on
which interval is chosen. The relevant seminorms are equivalent. The 
arguments of \cite{ringstrom06a} show that the set ${\cal G}_c$ is open in 
the $C^\infty$ topology. For the usual Gowdy case it is shown in 
\cite{ringstromta} that  ${\cal G}_c$ is also dense. It has not been 
verified whether the analogous statement is true for twisted Gowdy solutions.

\section{Bianchi type VII${}_0$ and the circular loop spacetimes}
\label{bianchi7}

This section is concerned with the circular loop spacetimes (see 
\cite{chrusciel91}, Appendix B)
and their relation to spacetimes of Bianchi type VII${}_0$. In contrast to
the spacetimes of Bianchi types II and VI${}_0$ considered elsewhere in 
this paper there is no topological twist in this case. The spatial topology
is $T^3$ but there is a geometrical twist. In terms of the wave map 
formulation the solutions to be considered here are again equivariant with
respect to a Killing vector of the hyperbolic plane. This Killing vector
looks complicated when expressed in terms of the coordinates $P$ and $Q$
and so it is convenient at this point to convert to coordinates adapted to
the disc model of the hyperbolic plane. They are defined by the relations 
that $\Phi\cos\Theta$ and $\Phi\sin\Theta$ are the real and imaginary parts 
of the complex quantity
\begin{equation}
\frac{Q+i(e^{-P}-1)}{Q+i(e^{-P}+1)}
\end{equation}
respectively. The image of the $(P,Q)$ plane under this mapping is the region
given by $0\le\Phi<1$. This is a polar coordinate system with origin at
$\Phi=0$.

The wave map equations take the form
\begin{eqnarray}
&&\Phi_{tt}+t^{-1}\Phi_t-\Phi_{\theta\theta}=\frac12\sinh2\Phi
(\Theta_t^2-\Theta_\theta^2),     \\
&&\sinh^2\Phi(\Theta_{tt}+t^{-1}\Theta_t-\Theta_{\theta\theta})=
\sinh 2\Phi(-\Phi_t\Theta_t+\Phi_\theta\Theta_\theta)
\end{eqnarray}
and the evolution equation for $\lambda$ is 
\begin{equation}
\lambda_t=t[(\Phi_t^2+\Phi_\theta^2)+\sinh^2\Phi(\Theta_t^2+\Theta_\theta^2)].
\end{equation} 
In these variables the metric on the group orbits is
\begin{equation}
e^{\bar\Phi}((\cos \frac12\alpha\theta) dx+(\sin\frac12\alpha\theta) dy)^2
+e^{-\bar\Phi}((-\sin \frac12\alpha\theta) dx+(\cos\frac12\alpha\theta) dy)^2.
\end{equation}
Let the initial data for $\Phi$, $\Phi_t$, $\Theta$ and $\Theta_t$ be denoted
by $\Phi_0$, $\Phi_1$, $\Theta_0$ and $\Theta_1$ respectively. The circular
loop spacetimes are defined by the conditions $\Phi(t,\theta)=\bar\Phi(t)$
and $\Theta(t,\theta)=\alpha\theta$ for a constant $\alpha\ne 0$ or by 
corresponding conditions on the initial data. The equation for $\Theta$ is 
satisfied identically by this ansatz while that for $\Phi$ reduces to
\begin{equation}\label{Phibar}
\frac{d^2\bar\Phi}{dt^2}+t^{-1}\frac{d\bar\Phi}{dt}=-\frac{\alpha^2}2
\sinh2\bar\Phi.
\end{equation}
The metric on the group orbits takes the form
\begin{equation}
e^{\bar\Phi}((\cos \frac12\alpha\theta) dx+(\sin\frac12\alpha\theta) dy)^2
+e^{-\bar\Phi}((-\sin \frac12\alpha\theta) dx+(\cos\frac12\alpha\theta) dy)^2
\end{equation} 
and the spacetime metric is of Bianchi type VII${}_0$. The wave map is 
equivariant with respect to
the Killing vector $\alpha\frac{\partial}{\partial\Theta}$. In this way it
can seen that the circular loop form defines the same class of solutions of
the vacuum Einstein equations as Bianchi type VII${}_0$. Notice also the
remarkable fact that with a change of notation (\ref{Phibar}) is identical
to (\ref{Vbar2}). The author has found no explanation for this coincidence
which means that the dynamics of solutions of Bianchi types VI${}_0$ and
VII${}_0$ are controlled by the same ODE.

In \cite{chrusciel91} the late-time behaviour of the circular loop spacetimes 
was determined. Among other things it was shown that $\bar\Phi$ and 
$\bar \Phi_t$ are $O(t^{-\frac12})$ as $t\to\infty$, that $tH(t)$ converges to a 
constant $H_\infty$ as $t\to\infty$ and that the spacetimes are future 
geodesically complete. The constant $H_\infty$ is strictly positive for any 
circular loop
spacetime. (Bianchi type I solutions are not considered to belong to this 
class.) All these statements were later extended to general Gowdy spacetimes
in \cite{ringstrom04}. Independently of this the late-time behaviour of
vacuum spacetimes of Bianchi type VII${}_0$ was analysed in \cite{ringstrom01}.
Knowing the relation between the Bianchi type VII${}_0$ and the circular loop
spacetimes, those results in \cite{ringstrom01} which concern solutions of
Bianchi type VII${}_0$ can easily be deduced from the results of 
\cite{chrusciel91}. In \cite{ringstrom01} it was shown that two of the 
Wainwright-Hsu variables $N_1$ and $N_2$ tend to the same constant value.
It turns out that this constant is equal to $12H_\infty^{-1}$.

\section{Late time dynamics}\label{latetime}

An important property of Gowdy models is the existence of a functional,
often called energy, whose dependence on time is monotone. It is given by
\begin{equation}
H=\frac12\int_0^{2\pi}P_t^2+P_\theta^2+e^{2P}(Q_t^2+Q_\theta^2) d\theta.
\end{equation}
In the Bianchi type II case the same quantity is monotone non-increasing.
The proof is essentially the same. It is just necessary to check that no
additional boundary terms arise during partial integration. The relevant
identity is
\begin{equation}\label{hdot}
\frac{dH}{dt}=-t^{-1}\int_0^{2\pi} P_t^2+e^{2P}Q_t^2 d\theta
+[P_tP_\theta+e^{2P}Q_tQ_\theta]_0^{2\pi}.
\end{equation}
Since $P$, $P_t$, $Q_t$ and $Q_\theta$ are periodic in the twisted type II 
case the boundary term vanishes. The boundary term also vanishes in the 
twisted type VI${}_0$ case. It should be noted that in both cases the
energy density is periodic so that the apparently arbitrary choice of 
$\theta=0$ as the starting point of integration has no effect on the 
value of the integral. An analogous definition using a different starting 
point gives the same answer. 

In Gowdy models it has been proved that $H(t)=O(t^{-1})$ as $t\to\infty$ 
\cite{ringstrom04}. This is done in two steps. First, it is shown that
if the energy is ever smaller than a certain threshold $H_1$ then it is 
$O(t^{-1})$ as $t\to\infty$. Second, it is shown that the energy tends to 
zero for all solutions. Twisted Gowdy solutions of type VI${}_0$ satisfy the 
inequality $H(t)\ge\pi\alpha^2$ and so in that case this strategy must be 
modified if it is to have a chance of success. It may be conjectured that
the energy tends to zero as $t\to\infty$ for twisted Gowdy solutions of
type II and to $\pi\alpha^2$ for solutions of type VI${}_0$. In the 
homogeneous case this follows from known results. It is also not difficult
to treat the homogeneous case directly, as will now be shown.

In the type II case, suppose first that $\bar P_t(t_1)>0$ for some $t_1$. Then
from (\ref{gowdyPhom}) for $t\ge t_1$
\begin{equation}
(t\bar P_t)_t(t)\le -\alpha^2 t_1e^{2\bar P(t_1)}
\end{equation} 
as long as $\bar P_t$ stays positive. It follows that $\bar P_t$ must become 
zero after a finite time. When $\bar P_t$ is zero its derivative is negative. 
This means that once it reaches zero $\bar P_t$ can never become positive 
again. Moreover it must become negative immediately after the time when it is 
zero. Thus to study the late time behaviour it may be assumed without loss of
generality that $\bar P_t$ is always negative. In particular $\bar P$ tends to 
a limit $\bar P^\infty$, finite or infinite, as $t\to\infty$. Suppose now that
$\bar P^\infty>-\infty$. Then
\begin{equation}
(t\bar P_t)_t(t)\le -\alpha^2 e^{2\bar P^\infty}t.
\end{equation}
Integrating this twice shows that $\bar P\to -\infty$ as $t\to\infty$, 
contradicting the assumption on $\bar P^\infty$. Thus in fact 
$\bar P(t)\to -\infty$ as $t\to\infty$. Now $H$ tends to a limit $H_0\ge 0$ 
as $t\to\infty$ and it follows that $\lim_{t\to\infty}\bar P_t=-\sqrt{2H_0}$. 
If $H_0$ were positive then $-\bar P$ would grow at least linearly and 
$e^{2\bar P}$ would decay at least exponentially.  This would imply that 
$t\bar P_t$ tends to a finite limit, a contradiction. Hence $H_0=0$.

In the type VI${}_0$ case the boundedness of the energy shows that $\bar V$ 
and $\bar V_t$ are bounded.  Using the evolution equation (\ref{Vbar2}) this 
implies the boundedness of $\bar V_{tt}$. This allows the use of a compactness
argument. Let $\{t_n\}$ be a sequence of times tending to infinity. Define
translated quantities by $\bar V^n(t)=\bar V(t-t_n)$. Then $\bar V^n$ and 
$\bar V^n_t$ satisfy uniform $C^1$ bounds on any compact time interval. By the 
Arzela-Ascoli theorem \cite{rudin} it follows that, possibly after passing to a 
subsequence, $\bar V^n$ converges uniformly on finite time intervals to
a limit $\bar V^*$ which satisfies the equation 
$\bar V^*_{tt}=-\frac12\alpha^2\sinh 2\bar V^*$. If the limit $H_0$ of $H$ as 
$t\to\infty$ is greater than $\pi\alpha^2$ then the limiting solution is
not identically zero. Moreover it is straightforward to show that 
it is periodic. Thus there exists $T>0$ such that 
$\bar V^*(t+T)=\bar V^*(t)$ for all $t$. Let 
$\eta=\int_{t_0}^{t_0+T}(\bar V^*_t(t))^2 dt$. Then
\begin{equation} 
\int_{t_0}^{t_0+mT}t^{-1}(\bar V^*_t(t))^2 dt
\ge\eta\sum_{k=1}^m(t_0+kT)^{-1}
\end{equation}
Since the sum on the right hand side diverges as $m\to\infty$ there exists
an integer $M$ such that
\begin{equation}
\int_{t_0}^{t_0+MT}t^{-1}(\bar V^*_t(t))^2 dt\ge 2(H(t_0)-H_0).
\end{equation}
Using the uniform convergence of the sequence $\bar V^n$ to its limit 
$\bar V^*$ on the interval $[t_0,t_0+MT]$ shows that 
\begin{equation}
\int_{t_0}^\infty t^{-1}(\bar V^n_t(t))^2 dt >(H(t_0)-H_0)
\end{equation} 
for $n$ sufficiently large and this contradicts the fact that $H$ is always 
positive. Thus in fact $\bar V^*$ is identically zero. From this it can be 
concluded that $H\to \pi\alpha^2$ as $t\to\infty$. Recall that the asymptotics
of vacuum solutions of types VI${}_0$ and VII${}_0$ are governed by the same
basic equation. It would be interesting to do a detailed comparison 
between the results in the literature of relevance to the detailed asymptotics
of solutions of this equation. Apart from the papers \cite{chrusciel91} and
\cite{ringstrom01} mentioned in the last section there is the work of 
Heinzle and Ringstr\"om \cite{heinzle09} on the late-time behaviour of 
solutions of Bianchi type VI${}_0$ based on the Wainwright-Hsu system
\cite{wainwright89}. Due to the variety of formulations of the equations
and notations used in the different papers a comparison of this type 
would involve heavy computations. 

A relatively simple case to start with in studying the behaviour of the energy
as $t\to\infty$ in inhomogeneous spacetimes is that of polarized twisted Gowdy 
solutions of type  VI${}_0$ since there the main field equation is linear. Let 
$P(t,\theta)$ be a solution of this type. Let $\tilde P$ be the explicit 
solution defined by $\alpha\theta$. Due to the linearity of the equation for 
$P$ the difference $P-\tilde P$ is also a solution. Moreover $P-\tilde P$ 
is periodic in $\theta$ and so is an ordinary polarized solution. Hence its 
asymptotics can be deduced from results proved in \cite{jurke}. It follows in 
particular that the energy of any polarized Gowdy solution of type VI${}_0$ 
tends to $\pi\alpha^2$ as $t\to\infty$. The energy $\tilde H$ of a twisted 
solution of type II obtained from a Gowdy solution by the Gowdy-to-Ernst 
transformation need not be equal to the energy $H$ of the original solution 
but it does satisfy $\tilde H(t)=H(t)+O(t^{-2})$. It follows from the known 
results on Gowdy solutions that $\tilde H(t)=O(t^{-1})$ for twisted type II 
solutions obtained in this way. Note that there are ordinary Gowdy solutions 
which have $H=0$. They are isometric to the Kasner solution with Kasner 
exponents $(-\frac13,\frac23,\frac23)$. A twisted Gowdy solution of type II 
never has $H=0$  and this applies in particular to the homogeneous solutions 
obtained by transforming the solutions of type I which satisfy $H=0$. 

It is worth noting that the results on the dynamics of Gowdy solutions 
proved by Ringstr\"om are results on solutions of the Gowdy equations and
are not dependent on the restriction (\ref{lambdaint}) arising from the 
equation for $\lambda$. In this context it is interesting to remark that there 
is an explicit class of solutions of the Gowdy equations whose dynamics is
easy to analyse but which almost always violate (\ref{lambdaint}). In these
solutions $P(t,\theta)=-\frac12\log t$ and $Q(t,\theta)=q(t-\theta)$ for 
an arbitrary function $q$ of one variable. They are characterized by the 
fact that they are fixed points of the Gowdy-to-Ernst transformation.
If $q$ is periodic these solutions have Gowdy symmetry while if 
$q(x+2\pi)=q(x)+2\pi\alpha$ they have twisted Gowdy symmetry of type II. They 
never satisfy (\ref{lambdaint}) unless $q$ is constant. The known results on 
the decay of solutions of the Gowdy equations apply to these solutions and the 
fact that $H(t)=O(t^{-1})$ can be read off directly in this case. In addition 
it is seen that solutions of this kind with type II symmetry have the same 
decay of $H$. The asymptotics of 
these spacetimes near the singularity can also be read off directly. In
the notation of equations (12) and (13) of \cite{ringstrom06a} the function 
$q$ of that paper coincides with that used here, the function $\phi$ vanishes 
identically, $v_a(\theta)=\frac12$ and $\psi(\theta)=q'(\theta)$. It does
not seem that solutions of the Gowdy equations of this kind can be 
interpreted as coming from spatially compact spacetimes. These solutions
belong to a class discussed by Wainwright and Marshman \cite{wainwright79}
\footnote{I thank Woei-Chet Lim for drawing my attention to this}.
To see the relation set $m=-\frac{3}{16}$ (which is the vacuum condition)
and $q=1$ in Case I of \cite{wainwright79}.

Some evidence has now been collected that in twisted models of type II and 
VI${}_0$ the energy tends to the limits zero and $\pi\alpha^2$ respectively
as $t\to\infty$. Unfortunately this has not yet been proved in general,
even in the case that the energy is initially small. The proofs of these
statements for the usual Gowdy spacetimes make extensive use of the averages
of the unknowns in space. In the twisted models global averages do not always
make sense but it is possible to define analogous quantities by the formula
\begin{equation}
\langle f(t,\theta)\rangle=\frac{1}{2\pi}\int_{\theta-\pi}^{\theta+\pi}
f(t,\sigma)d\sigma.
\end{equation}
If $f$ is periodic then $\langle f\rangle$ is equal to the average value of 
$f$ and, in particular, independent of $\theta$. For the functions $P$ and $Q$ 
in twisted models this is in general no longer the case. The averaged 
quantities satisfy the same boundary conditions as the original ones. This 
means in particular that in type II the difference $Q-\langle Q\rangle$ is 
periodic with integral zero. We also have the identity 
$\langle Q\rangle_\theta=\alpha$. The starting point for the proof of 
the asymptotics in the case of initial data with small energy is a
differential inequality for a suitable corrected energy of the form
$H+\Gamma^P+\Gamma^Q$ where
\begin{eqnarray}
&&\Gamma^P=\frac{1}{2t}\int_0^{2\pi}(P-\langle P)P_t d\theta,\\
&&\Gamma^Q=\frac{1}{2t}\int_0^{2\pi}e^{2\langle P\rangle}(Q-\langle Q\rangle)Q_t
d\theta.
\end{eqnarray}
The integrands in these formulas are periodic in both types II and VI${}_0$.
Many of the calculations which lead to the important differential inequality 
work just as well in the twisted type II case but there is one problematic 
term which is left over. This is of the form 
$\int_0^{2\pi}e^{2\langle P\rangle}
\langle Q(t,\theta)\rangle_\theta Q_\theta(t,\theta)d\theta.$ 
For an ordinary Gowdy solution it is zero but it the type II twisted
case it is equal to $\alpha^2e^{2\langle P\rangle}$. It is difficult to see how 
the latter expression could be estimated in way which would
lead to a useful differential inequality similar to that obtained in 
the ordinary Gowdy case. Thus this technique of proof seems to fail in the
type II case.

Since $P$ is periodic in type II it is possible to show that 
$|P-\langle P\rangle|\le CH^{\frac12}$ for a constant $C$ which only depends 
on the initial vakue of $H$ and that 
\begin{equation}
C^{-1}e^P\le e^{\langle P\rangle}\le Ce^P.
\end{equation}
Using this it can be shown that
\begin{equation}
e^{2\langle P\rangle}\le CH\alpha^{-2} 
\end{equation}
It follows that there is one important way in which the type II twisted case 
differs from the untwisted case: if $H$ tends to zero in a type II solution 
as $t\to\infty$ then $\langle P\rangle$ tends to $-\infty$ in that limit. For 
ordinary Gowdy solutions, on the other hand, there are solutions where 
$\langle P\rangle$ tends to $-\infty$ but also solutions where it tends to 
$+\infty$ and solutions where it remains bounded for all time. If  
$\langle P\rangle$ tends to $-\infty$ then $P$ tends uniformly to
$-\infty$.

\section{Comparison with linearized perturbations}\label{linearized}

Linearizing the full vacuum Einstein equations about the background given 
by a solution of Bianchi type II gives rise to a perturbation problem
which has been studied by Tanimoto \cite{tanimoto03}, \cite{tanimoto04}. 
In fact these papers deal mainly with the model problems where the linearized
Einstein equations are replaced by a scalar wave equation or the Maxwell
equations on the given background. If Tanimoto's work is specialized to the 
case with a symmetry corresponding to perturbations belonging to the class of
twisted Gowdy solutions then it should be possible to compare the result with 
the full nonlinear theory developed in Section \ref{latetime}. The analysis
of \cite{tanimoto03} and \cite{tanimoto04} is based on thinking of the 
spatial manifold of the Bianchi type II solution as a circle bundle over $T^2$
rather than a $T^2$ bundle over the circle. In the latter interpretation 
$\theta$ is a coordinate on the base manifold and $x$ and $y$ are coordinates
on the fibres. In the former $\theta$ and $y$ are coordinates on the base 
manifold while $x$ is a coordinate on the fibre. The case analysed in 
Section \ref{latetime} relates to perturbations which depend only on $\theta$ 
and, in particular, not on $x$. The coordinate $x$ in this paper corresponds 
to $z$ in \cite{tanimoto03}. Hence in Tanimoto's notation they satisfy the 
condition $m=0$. This case is not included in the theorems of \cite{tanimoto03} 
as a result of a genericity assumption. It is discussed in section 7 of 
\cite{tanimoto04}.   

With this motivation in mind, consider the dynamics of a solution of the 
wave equation $\nabla^\alpha\nabla_\alpha\psi=0$ on a background solution of 
Bianchi type II and, due to the subject of interest in this paper, restrict 
consideration to solutions which only depend on the coordinates $t$ and 
$\theta$. The wave equation on a spatially homogeneous spacetime takes the
form $\alpha^{-2}\psi_{TT}+(-\alpha^{-3}\alpha_T+\tr k)\psi_T=\Delta\psi$ with 
respect to a time coordinate $T$ which is constant on the hypersurfaces of
homogeneity and has lapse function $\alpha$. Consider for a moment the case
of a Kasner solution expressed in terms of an areal time coordinate. The 
wave equation takes the form
\begin{equation}\label{scalar}
\psi_{tt}+t^{-1}\psi_t=\psi_{\theta\theta}
\end{equation}
for any Kasner solution. This is just the polarized Gowdy equation. Since 
the wave equation is the simplest example of a wave map it is not surprising
that this coincidence is related to the representation of the Gowdy equations
in terms of a wave map with the domain metric (\ref{auxiliary}). In fact more
is true. In any Gowdy or twisted Gowdy spacetime the wave equation for a 
function $\psi$ depending only on $t$ and $\theta$ takes the form
(\ref{scalar}). In
particular this statement holds for any spatially homogeneous solution of
Bianchi type II or VI${}_0$. The linearization of the Gowdy equations about 
a homogeneous solution of type II (or even type I) are more complicated.
They read
\begin{eqnarray}
&&\tilde P_{tt}+t^{-1}\tilde P_t=\tilde P_{\theta\theta}
-2\alpha^2e^{2\bar P}\tilde P,
\\
&&\tilde Q_{tt}+t^{-1}\tilde Q_t=\tilde Q_{\theta\theta}
-2\bar P_t\tilde Q_t+2\alpha\tilde P_\theta
\end{eqnarray} 
where $\tilde P$ and $\tilde Q$ are the linearized variables corresponding to
$P$ and $Q$.

\section{Conclusions and outlook} 

In this paper the dynamics of solutions of the Gowdy equations with
unconventional boundary conditions corresponding to topologically twisted
manifolds was studied. A central question concerns the late-time behaviour
of the energy functional $H(t)$. In the untwisted case $H(t)=O(t^{-1})$ as
$t\to\infty$ and this estimate is in general sharp. By analogy we 
conjecture that the quantity $H(t)$, which is known to be non-increasing,
tends to zero as $t\to\infty$ in the type II case and to $\pi\alpha^2$ in
the type VI${}_0$ case. These statements were proved in some special cases
including infinite dimensional families of solutions. Unfortunately a
general proof of these statements was not found. Trying to apply the 
techniques which were successful in the usual Gowdy case runs up against 
obstacles and it seems that some essentially new ideas are needed to make 
more progress on this question.

The results of this paper concern solutions of the vacuum Einstein equations.
If instead the Einstein-Maxwell equations are considered then interesting new
issues arise. Even if the metric quantities satisfy the standard periodic
boundary conditions there is a topological feature which can have an 
important effect on the dynamics. If we define a Maxwell field to be 
a field tensor which satisfies the Maxwell equations then this issue is
not visible. It becomes so if we think of the Maxwell tensor as the 
curvature of a connection of a circle bundle over the spacetime manifold.
An alternative approach is to ask whether the field can be derived from
a global smooth vector potential. When it can it is possible to extend
the techniques from the vacuum case to the Einstein-Maxwell case to prove that 
the natural energy functional tends to zero as $t\to\infty$ 
\cite{ringstrom06b}. In the work of \cite{nungesser} on strong cosmic 
censorship in solutions of the Einstein-Maxwell equations with polarized 
Gowdy symmetry the existence of a 
global potential satisfying periodic boundary conditions was assumed.
Nothing was proved about the case where no potential exists. Using the 
concepts of the present paper a result on this question can be obtained.
It was shown in \cite{nungesser} that in solutions of the Einstein-Maxwell
equations with polarized Gowdy symmetry the metric function $P$ and a
potential $\chi$ satisfy the polarized Gowdy equations. If instead of 
assuming that $\chi$ is periodic in $\theta$ it is assumed that it
satisfies the boundary condition $\chi(t,\theta+2\pi)=\chi(t,\theta)$ a
problem is obtained which is equivalent to the vacuum Gowdy case with twisted
type II symmetry. These Einstein-Maxwell solutions can be interpreted as
corresponding to a situation on the torus with a Maxwell field which does
not come from a potential. We intend to develop these ideas concerning the 
Einstein-Maxwell equations further in a separate publication.  

Another possible direction in which the results of this paper can be extended
is to go to solutions of the vacuum Einstein equations in higher dimensions.
Assume that a solution in $n+1$ dimensions has $n-1$ commuting Killing 
vectors and satisfies a suitable condition of reflection symmetry 
generalizing that defining the Gowdy class in four dimensions. The field 
equations can be written in a way closely analogous to that in four 
dimensions with the central equations defining a wave map with values in
a suitable target space. There is a natural energy functional. It has
been shown in \cite{barbos}, generalizing the proofs of \cite{ringstrom04}, 
that this energy tends to zero as $t\to\infty$ in any dimension. There are
many connections between homogeneous models and models of Gowdy type in
higher dimensions which remain to be explored. Some more remarks on this
subject can be found in section 5 of \cite{goedeke}.

\end{document}